\documentstyle[11pt,psfig,twoside]{article}

\begin{document}

\newcommand{\dd}{\,{\rm d}}
\newcommand{\ie}{{\it i.e.},\,}
\newcommand{\etal}{{\it et al.\ }}
\newcommand{\eg}{{\it e.g.},\,}
\newcommand{\cf}{{\it cf.\ }}
\newcommand{\vs}{{\it vs.\ }}
\newcommand{\zdot}{\makebox[0pt][l]{.}}
\newcommand{\up}[1]{\ifmmode^{\rm #1}\else$^{\rm #1}$\fi}
\newcommand{\dn}[1]{\ifmmode_{\rm #1}\else$_{\rm #1}$\fi}
\newcommand{\upd}{\up{d}}
\newcommand{\uph}{\up{h}}
\newcommand{\upm}{\up{m}}
\newcommand{\ups}{\up{s}}
\newcommand{\arcd}{\ifmmode^{\circ}\else$^{\circ}$\fi}
\newcommand{\arcm}{\ifmmode{'}\else$'$\fi}
\newcommand{\arcs}{\ifmmode{''}\else$''$\fi}
\newcommand{\MS}{{\rm M}\ifmmode_{\odot}\else$_{\odot}$\fi}
\newcommand{\RS}{{\rm R}\ifmmode_{\odot}\else$_{\odot}$\fi}
\newcommand{\LS}{{\rm L}\ifmmode_{\odot}\else$_{\odot}$\fi}

\newcommand{\Abstract}[2]{{\footnotesize\begin{center}ABSTRACT\end{center}
\vspace{1mm}\par#1\par
\noindent
{~}{\it #2}}}

\newcommand{\TabCap}[2]{\begin{center}\parbox[t]{#1}{\begin{center}
  \small {\spaceskip 2pt plus 1pt minus 1pt T a b l e}
  \refstepcounter{table}\thetable \\[2mm]
  \footnotesize #2 \end{center}}\end{center}}

\newcommand{\TableSep}[2]{\begin{table}[p]\vspace{#1}
\TabCap{#2}\end{table}}

\newcommand{\FigCap}[1]{\footnotesize\par\noindent Fig.\  %
  \refstepcounter{figure}\thefigure. #1\par}

\newcommand{\TableFont}{\footnotesize}
\newcommand{\TableFontIt}{\ttit}
\newcommand{\SetTableFont}[1]{\renewcommand{\TableFont}{#1}}

\newcommand{\MakeTable}[4]{\begin{table}[htb]\TabCap{#2}{#3}
  \begin{center} \TableFont \begin{tabular}{#1} #4 
  \end{tabular}\end{center}\end{table}}

\newcommand{\MakeTableSep}[4]{\begin{table}[p]\TabCap{#2}{#3}
  \begin{center} \TableFont \begin{tabular}{#1} #4 
  \end{tabular}\end{center}\end{table}}

\newenvironment{references}%
{
\footnotesize \frenchspacing
\renewcommand{\thesection}{}
\renewcommand{\in}{{\rm in }}
\renewcommand{\AA}{Astron.\ Astrophys.}
\newcommand{\AAS}{Astron.~Astrophys.~Suppl.~Ser.}
\newcommand{\ApJ}{Astrophys.\ J.}
\newcommand{\ApJS}{Astrophys.\ J.~Suppl.~Ser.}
\newcommand{\ApJL}{Astrophys.\ J.~Letters}
\newcommand{\AJ}{Astron.\ J.}
\newcommand{\IBVS}{IBVS}
\newcommand{\PASP}{P.A.S.P.}
\newcommand{\Acta}{Acta Astron.}
\newcommand{\MNRAS}{MNRAS}
\renewcommand{\and}{{\rm and }}
\section{{\rm REFERENCES}}
\sloppy \hyphenpenalty10000
\begin{list}{}{\leftmargin1cm\listparindent-1cm
\itemindent\listparindent\parsep0pt\itemsep0pt}}%
{\end{list}\vspace{2mm}}

\def\TYLDA{~}
\newlength{\DW}
\settowidth{\DW}{0}
\newcommand{\dw}{\hspace{\DW}}

\newcommand{\refitem}[5]{\item[]{#1} #2%
\def\REFARG{#3}\ifx\REFARG\TYLDA\else, {\it#3}\fi
\def\REFARG{#4}\ifx\REFARG\TYLDA\else, {\bf#4}\fi
\def\REFARG{#5}\ifx\REFARG\TYLDA\else, {#5}\fi.}

\newcommand{\Section}[1]{\section{#1}}
\newcommand{\Subsection}[1]{\subsection{#1}}
\newcommand{\Acknow}[1]{\par\vspace{5mm}{\bf Acknowledgments.} #1}
\pagestyle{myheadings}

\def\thefootnote{\fnsymbol{footnote}}
%%%%%%%%%%%%%%%%%%%%%%%%%%%%%%%%%%%%%%%%%%%%%%%%%%%%%%%%%%%%%%%%%%%%%%%%%%%%

\begin{center}
{\Large\bf Period Changes of the LMC Cepheids Determined from the
Harvard, OGLE and ASAS Data}
\vskip1cm
{\bf
P~a~w~e~{\/l}~~P~i~e~t~r~u~k~o~w~i~c~z}
\vskip3mm
{Warsaw University Observatory, Al. Ujazdowskie 4, 00-478 Warszawa, Poland\\
e-mail: pietruk@astrouw.edu.pl}
\end{center}

\Abstract{
Observations of Cepheids in the Large Magellanic Cloud, made
over the last several decades, allow us to search for evolutionary
period changes. None of the Cepheid from our sample of 378 stars 
stopped pulsating. Also none of them showed a large period change
which could indicate mode switching. However for Cepheids with 
$ {\rm \log P > 0.9 } $ we found significant period changes,
positive as well as negative.  A comparison between the observed 
period changes and theoretical predictions shows moderate agreement
with some models (Bono et al. 2000), and a very large disagreement
with others (Alibert et al. 1999).  
The large differences between the models are likely caused by the
very high sensitivity of stellar evolution during core helium 
burning phase to even small changes in the input physics, as 
discovered by Lauterborn, Refsdal and Weigert (1971).
}{~}

{Galaxies: Magellanic Clouds -- Stars: Cepheids -- Stars: evolution}

\Section{Introduction}%1

The Period - Luminosity relation of the Cepheids is one of the most
fundamental tools for estimating distances in the Universe.  Various
observations of Cepheids provide tests for models of stellar structure,
evolution and pulsation. 

Cepheids are Population I stars undergoing core helium
burning.  They pulsate while crossing the instability strip
in the Hertzsprung--Russel diagram at the effective temperature 
${\rm \log T_{eff} \approx 3.8}$.  There is also a possibility
of observing a Cepheid during the first crossing of the strip, when
the star is in the Hertzsprung gap and evolves on a thermal time scale.

While a star crosses the instability strip its pulsation period
changes. This happens slowly even for massive stars. 
Hence a long time interval is needed to detect the changes. 
Some Cepheids in our Galaxy were observed for
almost 200 years.  The recent results of these studies were
published by Berdnikov and Ignatova (2000), who compiled the observations
of $ \delta $ Cep, $ \eta $ Aql and $ \zeta $ Gem, all showing
very strong period changes. Also Turner (1998) presented data on period
changes of 137 northern hemisphere Cepheids.  A quantitative relation
between the observed changes and those predicted by the evolutionary
models was investigated long time ago by Hofmeister (1967).
Recently Macri, Sasselov and Stanek (2001) reported on a dramatic 
change in the light curve of a Cepheid discovered by E. Hubble in M33. 
They suggest that the star stopped pulsating.

The Magellanic Clouds have very many known Cepheids, and a large data
set with their periods was published for the LMC by Payne-Gaposchkin 
(1971). These are the results of Harvard photographic observations 
made in the years: 1910 -- 1950.  Deasy and Wayman (1985) found 
that about 40 percent of a sample of 115 stars showed period variations,
apparently too rapid to be explained with the evolutionary models.
In the late 1990's a very large amount of CCD photometry for the Magellanic
Cloud Cepheids was obtained by several groups searching for gravitational 
microlensing.

The goal of this paper is to determine period changes in the LMC Cepheids
comparing the data published by Payne-Gaposhkin (1971) with the results
of two recent projects: OGLE (the Optical Gravitational Lensing Experiment,
Udalski et al. 1997), and ASAS (the All Sky Automated Survey, Pojma\'nski 
2000). We also compare the observed period changes with the predictions
of the recent stellar evolutionary models. 

\Section{Observational Data}%2

A digital version of the Payne-Gaposchkin (1971) data was kindly provided
to us by Dr. David Bersier.  All 1110 Cepheids have their HV (Harvard 
Variable) numbers.  The positions are given in H. Leavitt's coordinates. 
The moments of maxima correspond to the best observed epochs, and the periods
were estimated with the data spanning almost 50 years. There are also 
useful remarks about some stars, like a doubtful period or a light curve
with a large scatter, and the information about previously measured period. 
Contemporary data for the fainter Cepheids are taken from OGLE-II project 
(Udalski et al. 1999), while for the brightest stars they were obtained 
from the ASAS project (Pojma\'nski 2000). 

The OGLE and Harvard databases were matched using 2000.0 coordinates.
For each Cepheid from Harvard list, which should 
be in one of 21 OGLE fields, we looked for an OGLE Cepheid in 
a square 40'' on a side.  If there were more than one star in the
square, we chose that with a very close period.  We identified 368 stars
that way. Three Cepheids were not found.  HV 900 is too 
bright for OGLE camera (it is saturated).  HV 970 and HV 13032 are very 
faint in Harvard catalogue (average magnitudes are 16.84 
and 16.70 respectively) and the period is uncertain. There were some 
problems with HV 5651. Probably there is an error in the declination 
of this star in Payne-Gaposchkin's database.  The ASAS provided data 
for 11 brightest Cepheids.  We had a total of 379 stars for further 
analysis.

To be sure that stars were matched correctly we compared the
magnitudes (Fig. 1) and coordinates (Fig. 2) obtained from
the Harvard catalogue and from the OGLE or ASAS catalogs.  We rejected
the star HV 5761.  It is blended with a close companion.

We note that among the remaining 378 Cepheids 29 
are the first overtone pulsators and one is a double mode (fundamental and 
first overtone) pulsator. Cross-correlations of each variable and its 
parameters are available on the Internet at 
ftp://ftp.astrouw.edu.pl:/pub/pietruk/ceph.tab

\Section{Evolutionary models of Cepheids}%3

A recent theoretical survey of Cepheids' characteristics for
a number of evolutionary models was published by ABHA (Alibert,
Baraffe, Hauschildt, Allard 1999).  It contains parameters of 
stars at the blue and red edges of the instability strip for models in 
the ZAMS mass range ${\rm 2.75 - 12 M_\odot }$ with three chemical 
compositions: ${\rm (Z,Y) = (0.02,0.28)}$, ${\rm (0.01,0.25)}$ and 
$ {\rm (0.004,0.25) }$, representative of the Galaxy, LMC and SMC, 
respectively.  

Other theoretical models were recently published by Bono et al. (2000).
They adopted the same metallicities as ABHA but slightly different helium 
contents: ${\rm (Z,Y) = (0.02,0.27) }$, ${\rm (0.01,0.255) }$ and 
${\rm (0.004,0.23) }$.  Bono et al. gave the duration for every crossing
of the instability strip, but they gave the fundamental mode periods 
at the middle of the strip only.

There are some general properties of all evolutionary models.
For a given crossing of the instability strip the periods at 
the red edge are larger than those at the blue edge by an 
average factor 1.6.  The mean period, and the luminosity,
increase with the stellar mass. 
For a given chemical composition a star with a mass below certain value
has the evolutionary loop too small to enter the instability strip.  A 
star with a somewhat larger mass enters the instability strip, but the
maximum effective temperature reached in the loop phase is within the
strip, i.e. such stars enter and exit the instability strip through its
red edge.  Finally, still more massive stars cross the full strip twice
during their loop phase of evolution: first from the red to the blue, and
next they return from the blue to the red.

For these most massive stars there are three crossing through the 
instability strip: I, II, III, with the first referring to the very rapid
crossing of the Hertzsprung gap during the evolution following hydrogen
exhaustion in the core, and towards helium ignition in the core.
During this first crossing the pulsation period increases rapidly.
During the much slower loop phase the pulsation period decreases
in crossing II, and it increases again in crossing III, while the
star burns helium in the core and hydrogen in the shell.

Using the data from ABHA and Bono et al. tables we plotted in Fig. 3 
the crossing times as a function of period for ${\rm Z=0.01 }$.  It is clear
that models agree well for crossings I, but they disagree by up to
two orders of magnitude for the crossings II and III.  This is likely
the consequence of the phenomenon discovered decades ago
by Lauterborn, Refsdal and Weigert (1971), who found that stellar 
structure and evolution during the core helium burning phase is
very sensitive to even small changes in the input physics.

Since we have values of the pulsation periods $P_0$ and $P_1$ in
two moments of time $t_0$ and $t_1$ respectively (at the strip edges),
we define the theoretical rate of period change as
$$
r_{th} \equiv \frac{\Delta P}{\Delta t}\frac{1}{{P}^2}
=\frac{P_1-P_0}{t_1-t_0}\frac{1}{{P}^2}
\eqno(1)
$$
The scaling is chosen so that all model results of ABHA can be presented
in Fig. 4, which displays the rates of period change as a function of 
period for three metallicities.  The crossing I is well
separated from II and III, as the star is crossing the Hertzsprung
gap on a thermal time scale.

\Section{Comparison with the data}%4

We calculate the rate of observational period change using the equation 
$$
r_{obs} \equiv \frac{\Delta P}{\Delta t}\frac{1}{{P_1}^2}
=\frac{P_1-P_0}{t_1-t_0}\frac{1}{{P_1}^2}
\eqno(2)
$$
where $P_0$ is the old (Harvard) period at the moment of Cepheid light curve
maximum $t_0$, and $P_1$ is the new (OGLE or ASAS) period at the moment of 
maximum $t_1$.  We estimate the uncertainty of the rate of period change 
using the relation:
$$
\sigma_{obs} \approx \frac{\sigma _{P_1}}{t_1-t_0}\frac{1}{{P_1}^2}
\eqno(3)
$$ 
where $\sigma _{P_1}$ is the estimated error of the period as given by 
OGLE or ASAS.  Unfortunately, the error estimates of the Harvard periods,
$\sigma _{P_0}$, were not given by Payne-Gaposchkin.  Therefore,
$\sigma_{obs}$ is the lower bound of the observational error of the rate.
However, the periods determined from Harvard data are generally of high 
accuracy, as they are based on the observations covering several decades.
Hence, the real $\sigma_{obs}$ is not likely to be much larger than the
estimate given with the eq. (3).  We neglected the contribution of
${\rm t_0 }$ and ${\rm t_1 }$ uncertainties to the error balance.

The errors of OGLE periods were given as ${\rm \sigma _{P_1}=7 \times 10^{-5}
~ P_1 }$ by Udalski et al. (1999).  The corresponding errors for ASAS variables
were kindly calculated by Dr. Laurent Eyer using Hipparcos software.

Fig. 5 presents the ratio of the variance of the observed distribution 
of the measured period changes $\sigma _{dist}$ to the the average
nominal observational error 
$\sigma _{obs}$ defined with the eq. (3). We binned Cepheids into three
period groups.  Only the group with the longest periods, $ \log P > 0.9 $, 
has measurable period changes between the epoch of 
Harvard observations and the present observations of OGLE and ASAS.
Figs. 6 and 7, display a comparison between the
rates of period change as observed
for long period Cepheids, and the two sets of theoretical models, 
ABHA and Bono et al., respectively.  It is clear that some observed
rates are significant, i.e. much larger than their nominal errors,
and some are not significant.  It is also clear that the ABHA models 
for ${\rm Z = 0.01 }$ (corresponding to LMC Cepheids)
predict the rates of period change which are much larger than observed,
while the rates predicted by Bono et al. are comparable to the observed rates.

Bono et al. (2000) do not provide all the data we needed for Fig. 7.
We assumed that the period change, ${\rm \dot P }$, is constant during
the model crossing the instability strip. The lines
in Fig. 7 correspond to the variation of period in the denominator of
the formula ${\rm \dot P / P^2 }$ between the two edges of the instability
strip, assuming that the ratio of periods at the red to
the blue edges is 1.6.

\Section{Other evolutionary effects}

The OGLE photometry was done in B, V, and I bands.  Also,
the estimate of interstellar reddening was provided for each star.
Fig. 8 presents the color-period relation, where the ${\rm (V-I)_0 }$
index was corrected for the reddening (Udalski et al. 1999).
A slope of the instability strip, 
as well as its width are clearly apparent.
The models of ABHA predicted that the stellar evolution is
much faster near the red edge of the instability strip than
near the blue edge.  Therefore, we calculated ${\rm \delta (V-I)_0 }$,
which is the difference between the measured value of ${\rm (V-I)_0 }$,
and the value corresponding to the straight line drawn in Fig. 8 through
the middle of the instability strip.  The ABHA models predict that
there should be a correlation between the ${\rm \delta (V-I)_0 }$ 
parameter and the absolute value of the observed rate of period change,
${\rm | \dot P / P^2 | }$: the redder the star, the more rapid the
period change should be.  The observed diagram is shown in Fig. 9 for Cepheids
with measurable rate of period change, i.e. those with ${\rm \log P > 0.9 }$.
There is no apparent correlation.

\Section{Discussion}%5

There are several important conclusions following from our analysis.
None of 378 Cepheids has left the instability strip or
changed the pulsating mode during several decades separating
Harvard, OGLE and ASAS observations. This is consistent with the
probability for these processes.  We can estimate the probability of
leaving the strip in a time interval of up to 100 years.
The time it takes to cross the instability strip is approximately
given as ${\rm |P / \dot P | }$, which is observed to be $ \sim 3 \times 
10^5 $ years for ${\rm P = 10 }$ days, and $ \sim 3 \times 10^4 $ years
for ${\rm P = 100 }$ days (cf. Fig. 6 and 7).  Therefore, the probability
that a star with a period in the range $ 10 - 100 $ days (for which
${\rm \dot P }$ values are measurable) would get out of the instability
strip in just 100 years is only $ \sim 10^{-3} $.  For shorter period 
Cepheids the evolutionary time scales are longer and the corresponding 
probabilities are even smaller.

Deasy and Wayman (1985) noticed that the observed period changes were more
rapid than expected according to the models popular at that time.  We find
that the observed period changes are slower than predicted by the ABHA
models and about as rapid as expected by Bono et al. (2000).  Note:
in principle many factors may contribute to period changes (like mixing 
or He content), so the evolutionary predictions should provide a lower 
limit to what is observed.  Clearly, the predictions of the ABHA 
models cannot be right.

A histogram of Cepheid periods resulting from ABHA models predicts too few
long period Cepheids, or there are too many long period Cepheids observed
in the Magellanic Clouds.  This discrepancy, noted by ABHA, is a direct
consequence of the too rapid evolution of their models across the
instability strip during the loop phase leading to too rapid period
changes and too short lifetimes, and hence too few Cepheids.

It is surprising that we have not found any star undergoing the first
crossing of the instability strip.  The model evolutionary time scales 
corresponding to the first crossing are reliable, as these are
simply thermal time scales.  Indeed, the two sets of models agree
with each other (cf. triangles in Fig. 3).  The empirical crossing time 
scales for the loop phase are between $ \sim 3 \times 10^4 $ and 
$ 3 \times 10^5 $ years (previous paragraph), while the first 
crossing time scales are expected to be
$ \sim 4 \times 10^3 $ years for the long period Cepheids, i.e. only
a factor of $ \sim 25 $ shorter.  With the large number of observed
Cepheids we would expect to find some with large ${\rm \dot P }$ 
values, corresponding to the first crossing.  Yet, the most rapid 
period change observed is negative, i.e. 
it cannot correspond to the first crossing (note the point close to the
lower edge of Fig. 6 and Fig. 7, at ${\rm \log P \approx 1.3 }$).

The observational data could be considerably corrected.  If the Harvard
photometric data were available it would be possible to phase together
the old and the modern photometric measurements, improving considerably
the observational estimates of period changes, and providing more
reliable estimates of their errors.  However,
even with the results as presented in this paper there is a clear
need to refine theoretical models.

Note that neither ABHA nor Bono et al. (2000) models cover the longest
period Cepheids, i.e. the brightest and the most massive stars.  It would
be very useful to extend model calculations to masses large enough to
account for Cepheids with periods up to 130 days.

\Acknow{
I am grateful to Dr. K. Z. Stanek and Dr. J. Kalu\.zny for suggesting 
this project. I would like to thank Dr. D. Bersier for providing the paper
containing HV catalog, and the data in computer readable form.  I also thank
Dr. W. Dziembowski and Dr. G. Pojma\'nski for 
useful remarks and Dr. L. Eyer for calculating the errors of ASAS periods.
It is also a pleasure to acknowledge discussions with Dr. B. Paczy\'nski
during author's visit to Princeton University in July 2001, and for critical
reading of a draft of this paper. This work was supported in part
with the NSF grant AST--9820314 to B. Paczy\'nski.
}

%\newpage
% REFERENCES

\begin{figure}[htb]
\hglue-0.5cm\psfig{figure=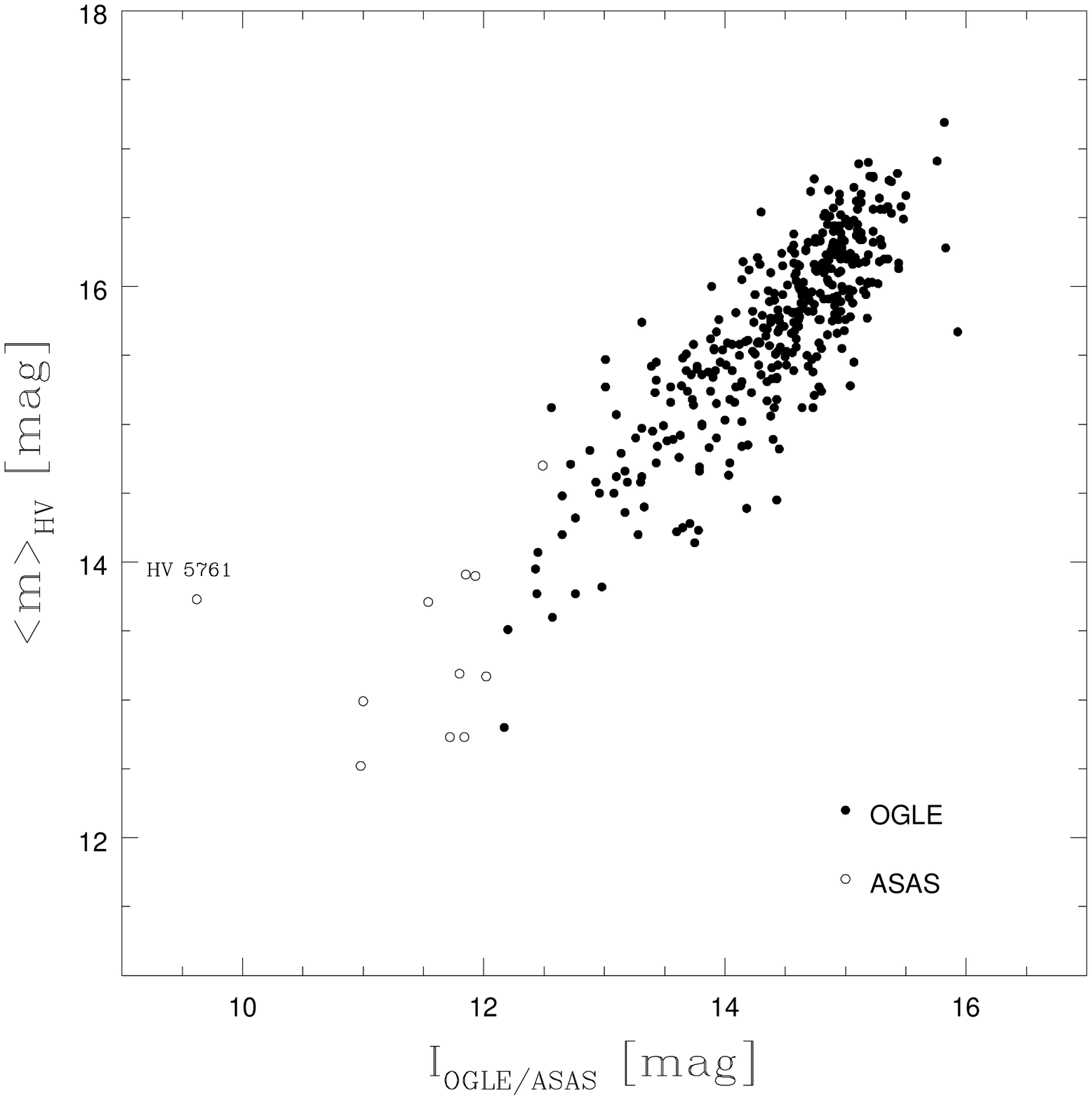,bbllx=0pt,bblly=0pt,bburx=590pt,bbury=720pt,width=12.5cm,
clip=,angle=0}
\vspace*{3pt}
\FigCap{
A comparison between the average Harvard magnitudes and I-band OGLE and ASAS
magnitudes for 379 Cepheids in the Large Magellanic Cloud.  Notice the
anomalous location of HV 5761.
}
\end{figure}

\begin{figure}[htb]
\hglue-0.5cm\psfig{figure=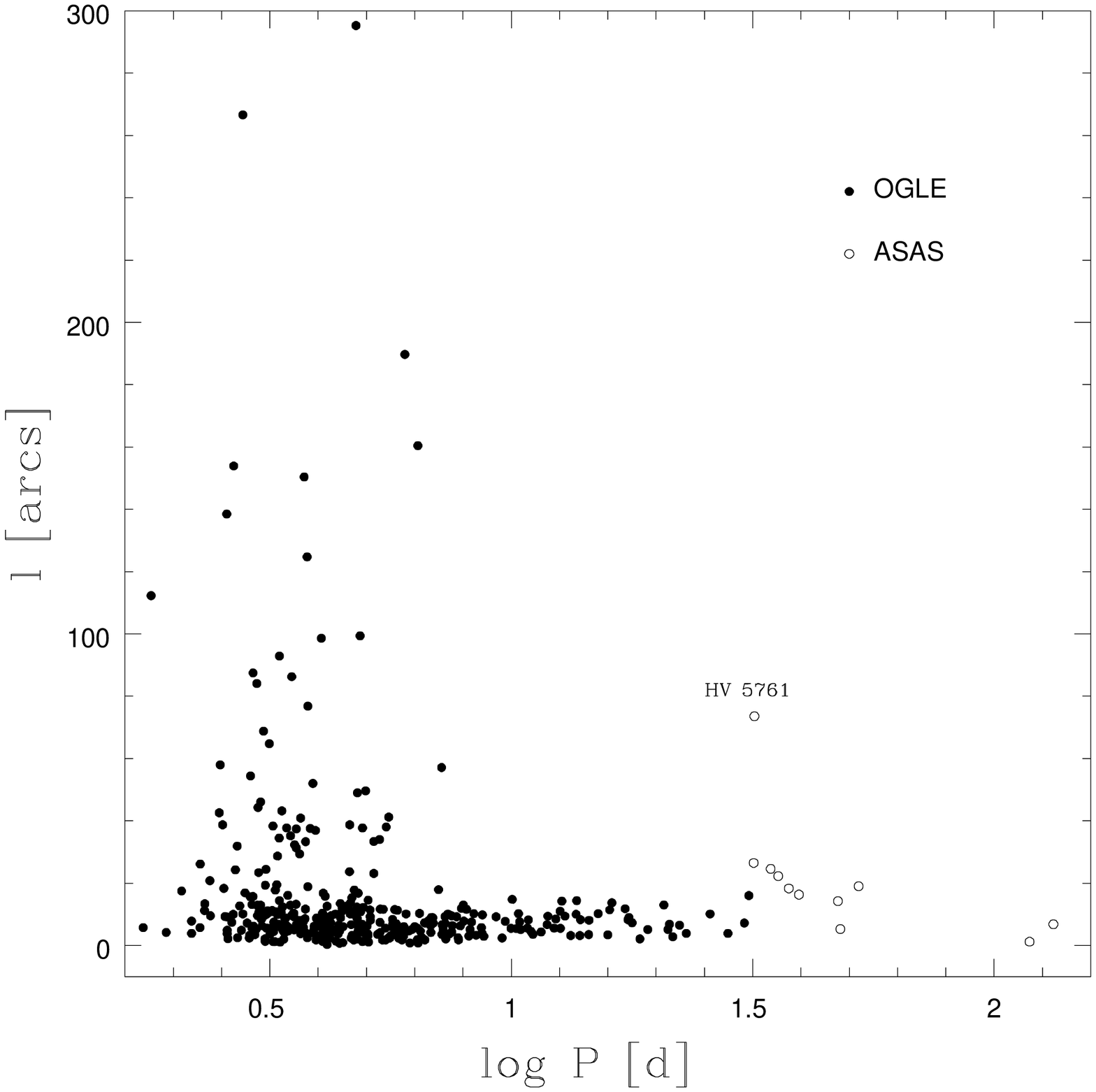,bbllx=0pt,bblly=0pt,bburx=590pt,bbury=720pt,width=12.5cm,
clip=,angle=0}
\vspace*{3pt}
\FigCap{
The difference in coordinates (in arcsecs) between the Harvard catalog
and the OGLE and ASAS catalogs is shown as a function of period.  The positions
are in a very good agreement for bright stars, with ${\rm \log P > 0.9 }$.
Notice the anomalous location of HV 5761.
}
\end{figure}

\begin{figure}[htb]
\hglue-0.5cm\psfig{figure=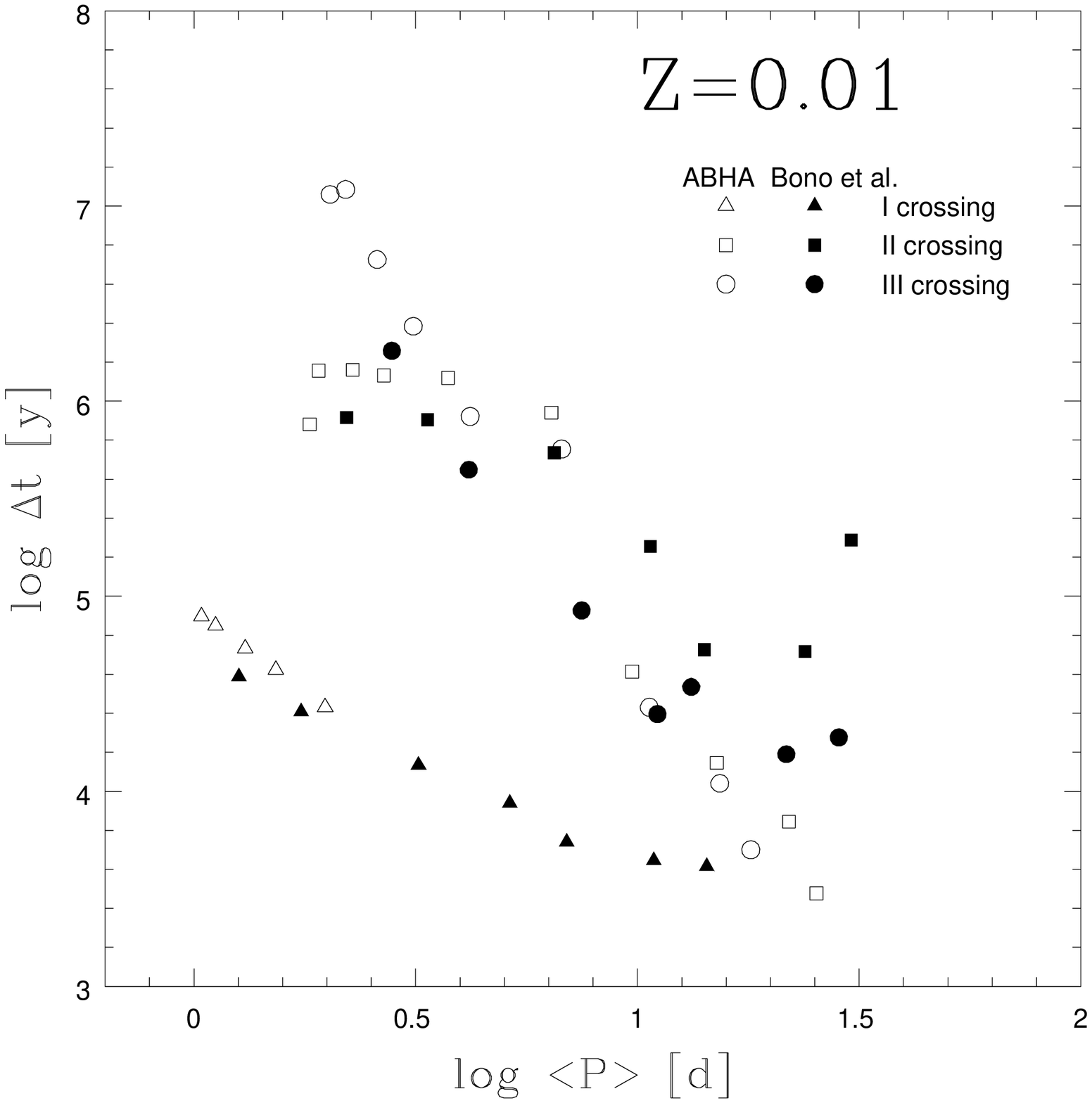,bbllx=0pt,bblly=0pt,bburx=590pt,bbury=720pt,width=12.5cm,
clip=,angle=0}
\vspace*{3pt}
\FigCap{
Crossing time of the instability strip as a function
of Cepheid period according to two theoretical paper: ABHA (Alibert, 
Baraffe, Hauschildt, Allard 1999) and Bono et al. (2000).  
Different symbols correspond to the 
three crossings of the instability strip. Notice that the two sets of
models agree well for the first crossing, which is fast, on a thermal time
scale. The agreement is very poor for the second and third crossings, which
are relatively slow, corresponding to the evolution in the loop phase.
}
\end{figure}

\begin{figure}[htb]
\hglue-0.5cm\psfig{figure=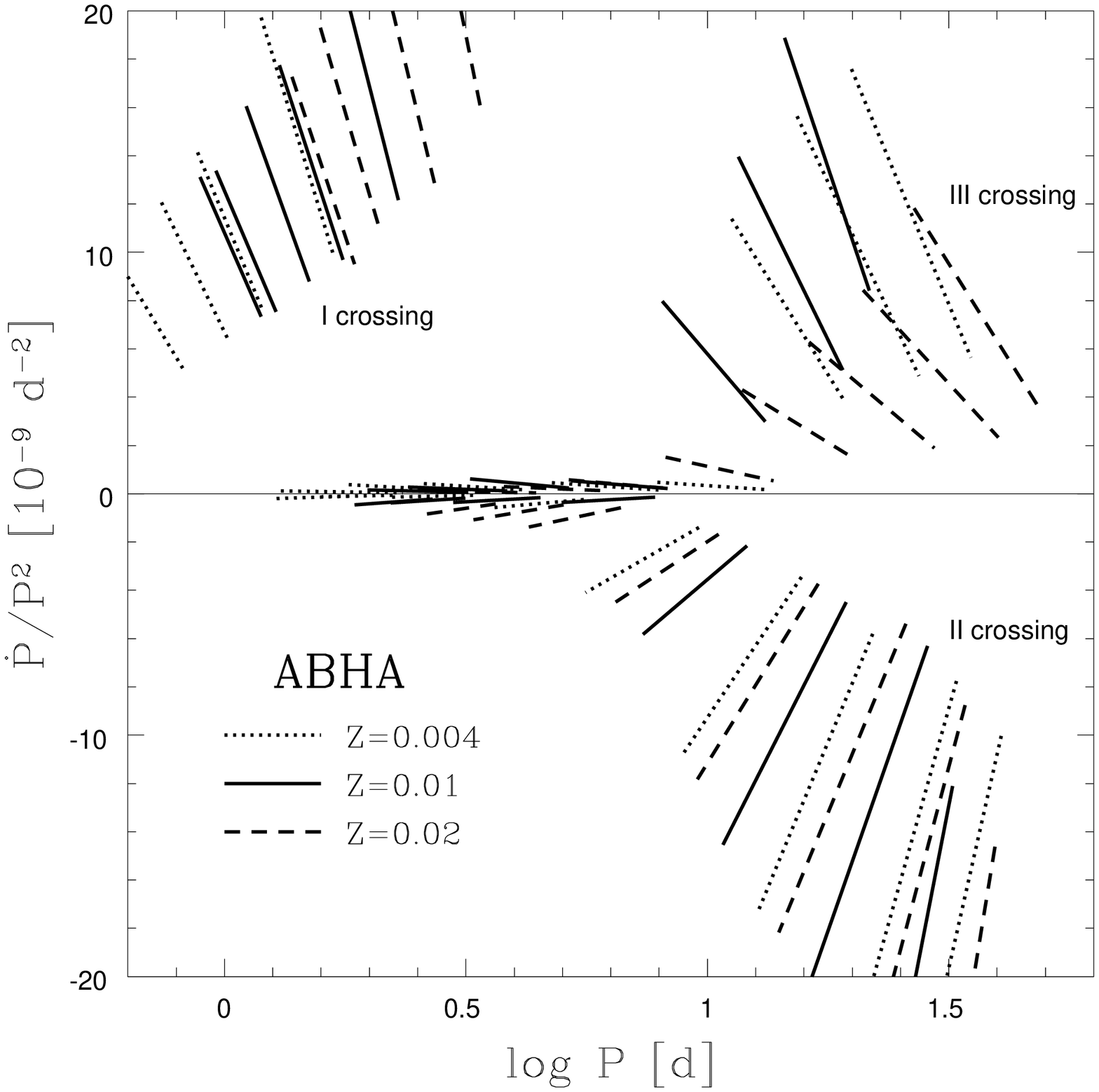,bbllx=0pt,bblly=0pt,bburx=590pt,bbury=720pt,width=12.5cm,
clip=,angle=0}
\vspace*{3pt}
\FigCap{
The rate of period changes predicted for the three crossings of Cepheid
instability strip is show as a function of period according to the models
calculated by ABHA (Alibert, Baraffe, Hauschildt, Allard 1999).
}
\end{figure}

\begin{figure}[htb]
\hglue-0.5cm\psfig{figure=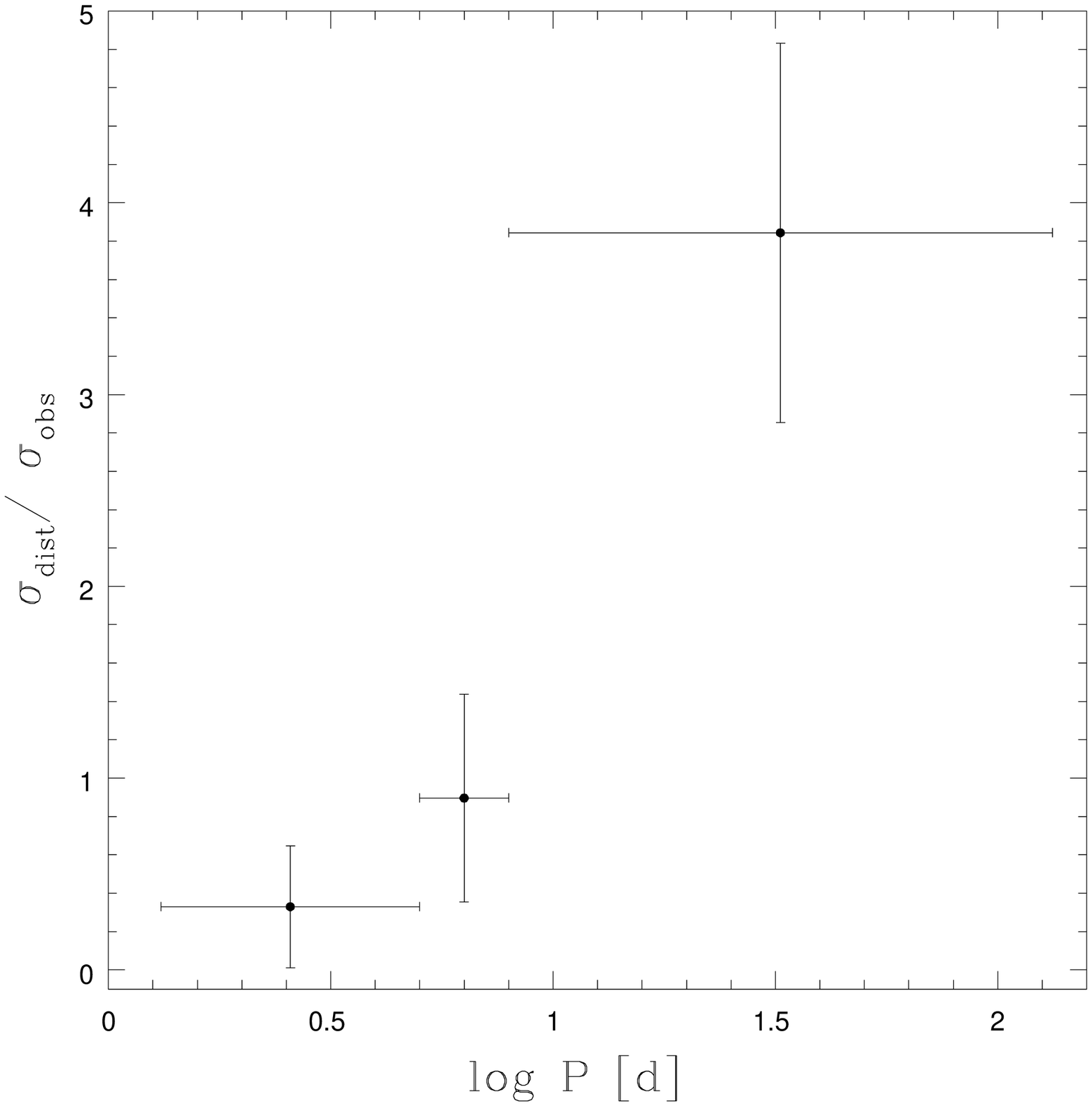,bbllx=0pt,bblly=0pt,bburx=590pt,bbury=720pt,width=12.5cm,
clip=,angle=0}
\vspace*{3pt}
\FigCap{
The ratio of the observed scatter in the measured period changes 
$ \sigma _{dist} $ to the nominal observational error $ \sigma _{obs} $
is shown for three groups of Cepheid periods.  It is clear that only
the group with the longest periods, ${\rm \log P > 0.9 }$, has measurable
period changes between the epoch of Harvard observations, and present
as observed by OGLE and ASAS.
}
\end{figure}

\begin{figure}[htb]
\hglue-0.5cm\psfig{figure=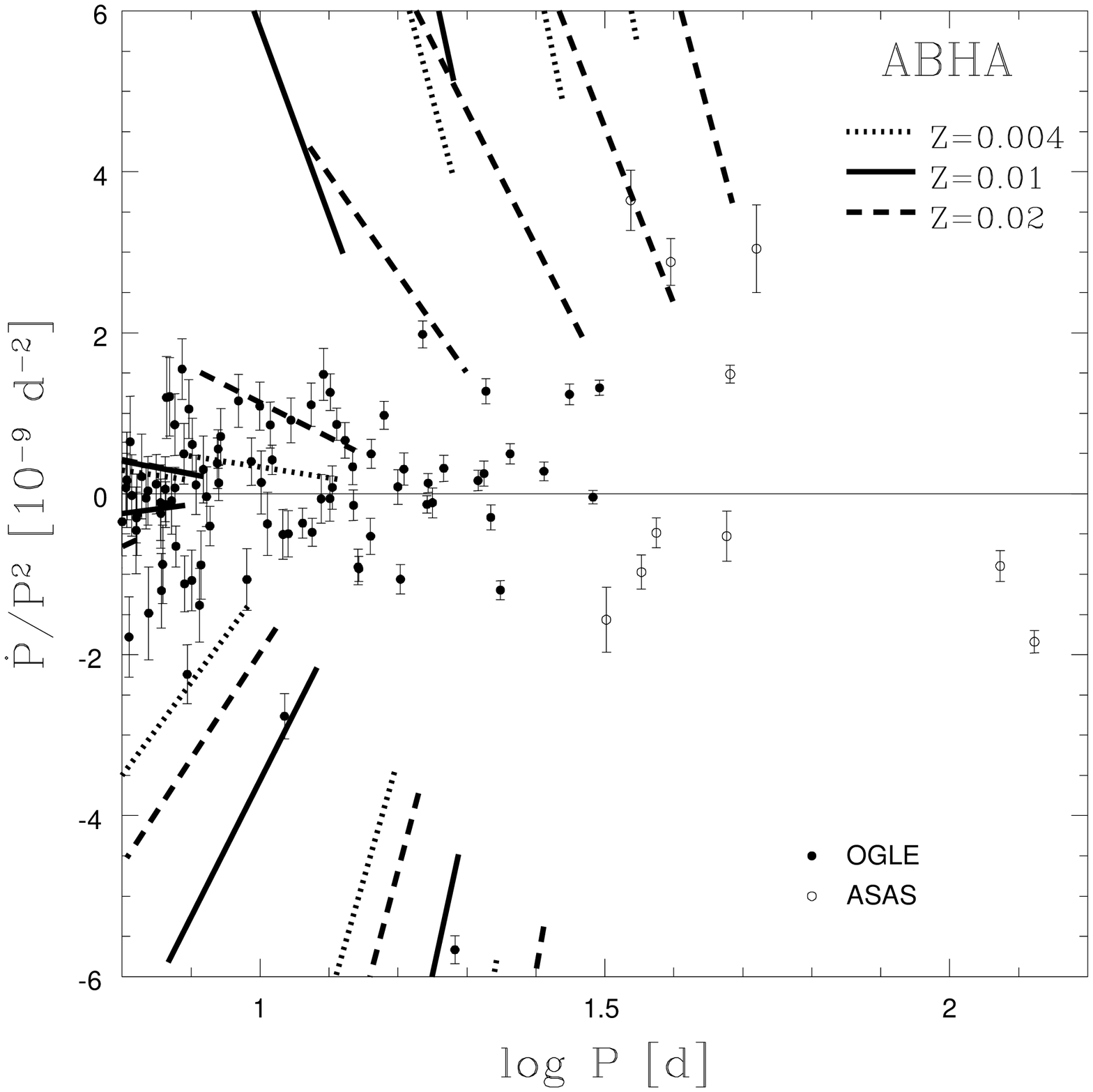,bbllx=0pt,bblly=0pt,bburx=590pt,bbury=720pt,width=12.5cm,
clip=,angle=0}
\vspace*{3pt}
\FigCap{
A comparison between the period changes predicted for Cepheids
with long periods by ABHA (Alibert, Baraffe,
Hauschildt, Allard 1999) and the Harvard, OGLE and ASAS observations.
The disagreement is striking.
}
\end{figure}

\begin{figure}[htb]
\hglue-0.5cm\psfig{figure=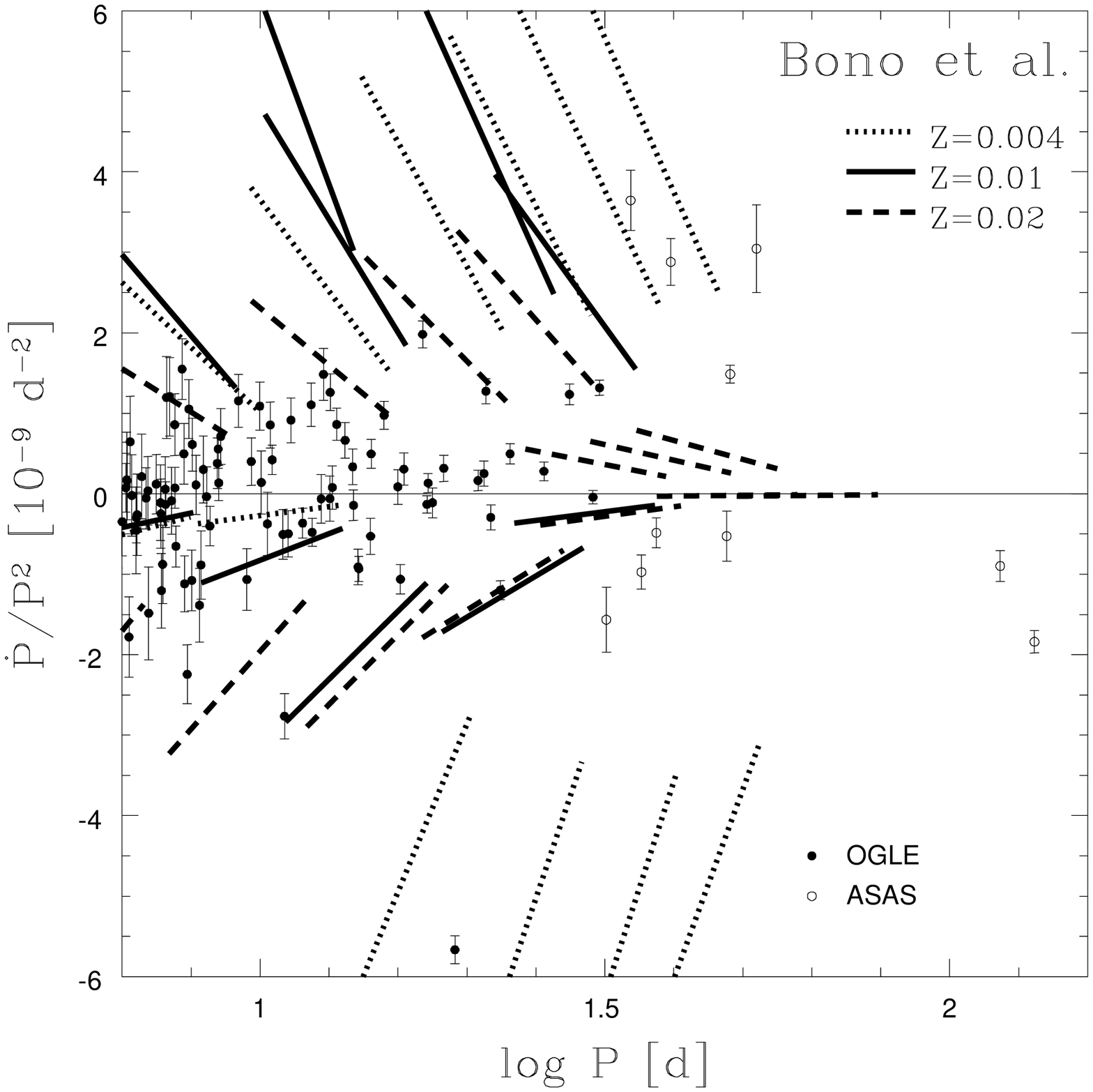,bbllx=0pt,bblly=0pt,bburx=590pt,bbury=720pt,width=12.5cm,
clip=,angle=0}
\vspace*{3pt}
\FigCap{
A comparison between the period changes predicted for Cepheids
with long periods by Bono et al. (2000)
and the Harvard, OGLE and ASAS observations. The models and
the observations are in approximate agreement.
}
\end{figure}

\begin{figure}[htb]
\hglue-0.5cm\psfig{figure=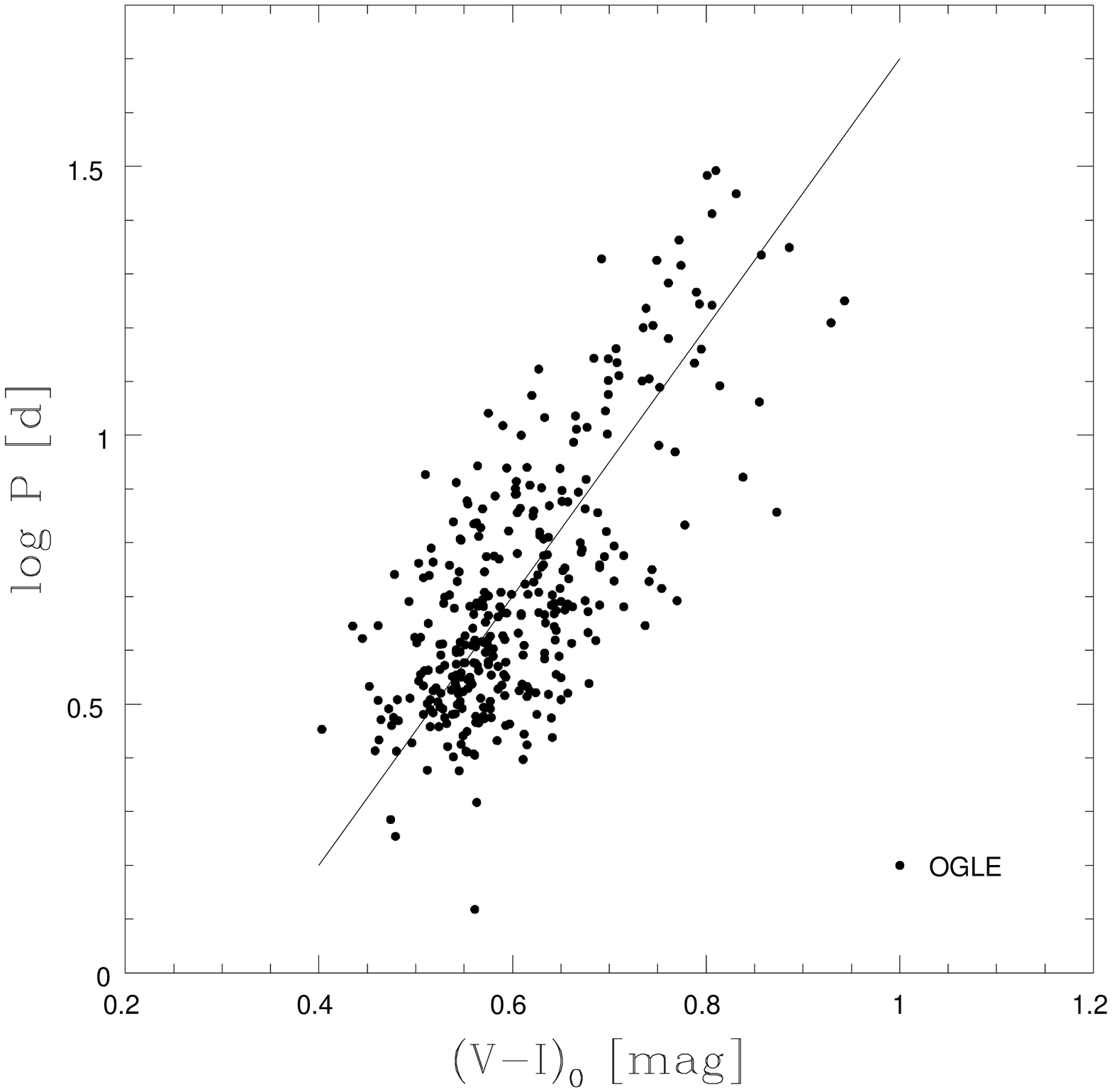,bbllx=0pt,bblly=0pt,bburx=590pt,bbury=720pt,width=12.5cm,
clip=,angle=0}
\vspace*{3pt}
\FigCap{
The observed color-period relation for the OGLE fundamental mode Cepheids.
}
\end{figure}

\begin{figure}[htb]
\hglue-0.5cm\psfig{figure=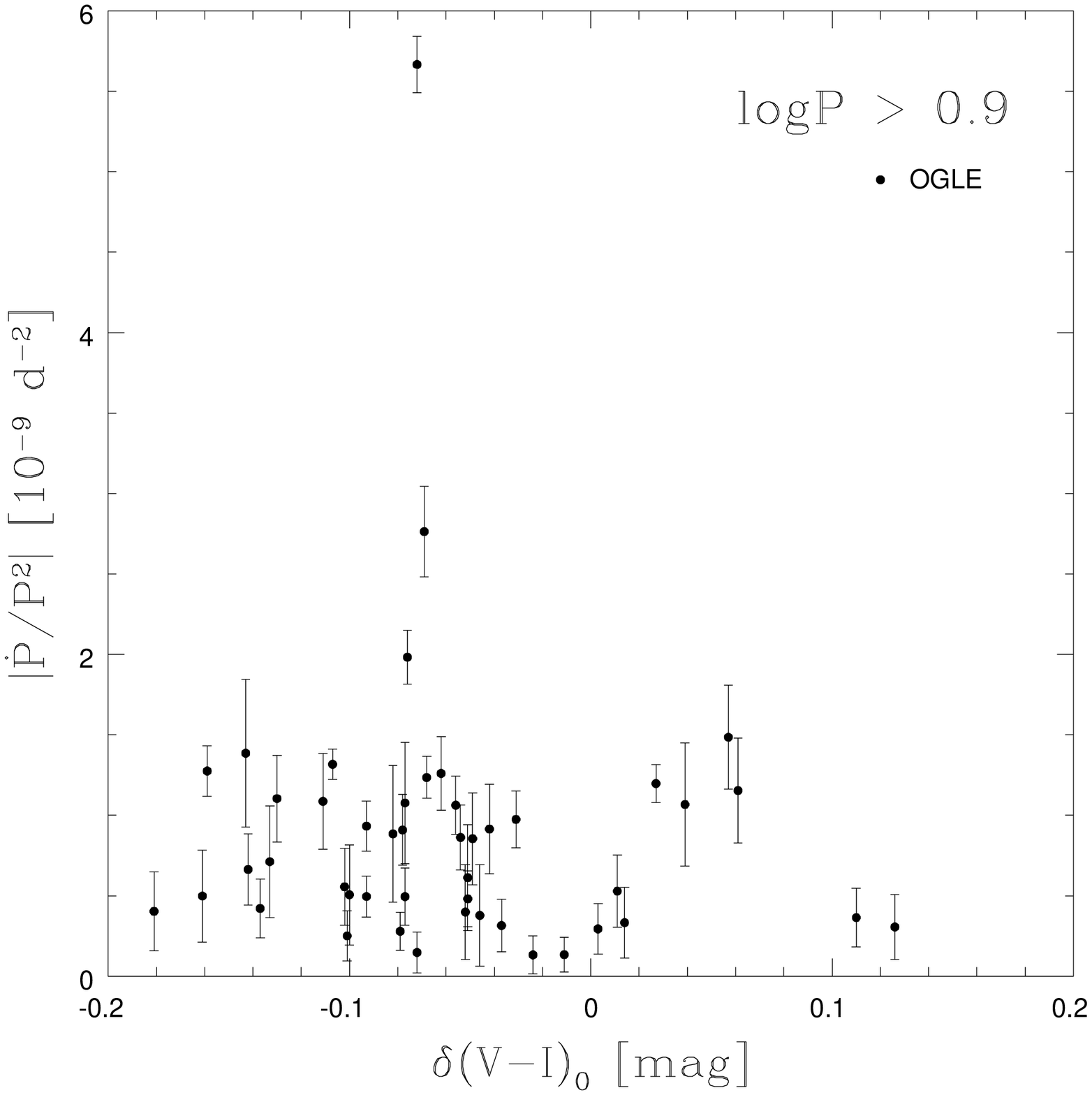,bbllx=0pt,bblly=0pt,bburx=590pt,bbury=720pt,width=12.5cm,
clip=,angle=0}
\vspace*{3pt}
\FigCap{
The dependence of the observed rate of Cepheid period change as a function
of color difference between the observed value $ {\rm (V-I)_0 } $ and the
line shown in Fig. 8.
}
\end{figure}


\begin{references}

\refitem{Alibert, C., Baraffe, I., Hauschildt, P., and Allard,F.}{1999}
   {\AA}{344}{551}
\refitem{Berdnikov, L. N., and Ignatova, V. V.}{2000}{New Astronomy}{4}{625}
\refitem{Bono, G., Caputo, F., Cassisi, S.,Marconi, M., Piersanti, L.,
   and Tornambe, A.}{2000}{\ApJ}{543}{955}
\refitem{Hofmeister, E.}{1967}{Z. Astrophys.}{65}{94}
\refitem{Deasy, H. P., and Wayman, A.}{1985}{\MNRAS}{212}{395}
\refitem{Lauterborn, D., Refsdal, S., and Weigert, A.}{1971}{\AA}{10}{97}
\refitem{Macri, L. M., Sasselov, D. D., Stanek, K. Z.}{2001}{\ApJ}{550}{L159}
\refitem{Payne-Gaposchkin, C. H.}{1971}{Smithsonian Contr. to Astroph.}{13}{}
\refitem{Pojma\'nski, G.}{2000}{\Acta}{50}{177}
\refitem{Turner, D.}{1998}{JAASVO}{26}{101}
\refitem{Udalski, A., Kubiak, M., and Szyma\'nski, M.}{1997}{\Acta}{47}{319}
\refitem{Udalski, A., Soszynski, I., Szyma\'nski, M., Kubiak, M., 
   Pietrzy\'nski, G., Wozniak, P., and Zebrun, K.}{2000}{\Acta}{50}{223}

\end{references}
\end{document}